# AGENT ENABLED MINING OF DISTRIBUTED PROTEIN DATA BANKS

G. S. Bhamra[1], A. K. Verma[2] and R. B. Patel[3]

[1]M. M. University, Mullana, Haryana, 133207 - India
[2]Thapar University, Patiala, Punjab, 147004- India
[3]Chandigarh College of Engineering & Technology, Chandigarh- 160019- India

## ABSTRACT

*Mining biological data is an emergent area at the intersection between bioinformatics and data mining (DM). The intelligent agent based model is a popular approach in constructing Distributed Data Mining (DDM) systems to address scalable mining over large scale distributed data. The nature of associations between different amino acids in proteins has also been a subject of great anxiety. There is a strong need to develop new models and exploit and analyze the available distributed biological data sources. In this study, we have designed and implemented a multi-agent system (MAS) called Agent enriched Quantitative Association Rules Mining for Amino Acids in distributed Protein Data Banks (AeQARM-AAPDB). Such globally strong association rules enhance understanding of protein composition and are desirable for synthesis of artificial proteins. A real protein data bank is used to validate the system.*

## KEYWORDS

*Knowledge Discovery, Association Rules, Intelligent Agents, Multi-Agent System, Bioinformatics.*

## 1. INTRODUCTION

Data Mining (DM) is a process to automatically extract some interesting and valid data patterns or trends representing knowledge, implicitly stored in large databases [1], [2]. Distributed Data Mining (DDM) is concerned with application of classical DM procedures in a distributed computing environment trying to make the best of the available resources. In DDM, DM takes place both locally at each geographically distributed site and at a global level where the local knowledge is merged in order to discover global knowledge. A DDM system is a very complex entity that is comprised of many components; mining algorithms; communication subsystems; resources management; task scheduling; user interface etc. It should provide efficient access to both distributed data and computing resources; monitor the entire mining procedure; and present results to users in appropriate formats. A successful DDM system is also flexible enough to adapt to various situations [3], [4], [5], [6], [7].

Intelligent software agent technology is an interdisciplinary technology. The motivating idea of this technology is the development and efficient utilization of autonomous software objects called agents, which have access to geographically distributed and heterogeneous information resources to simplify the complexities of distributed computing. They are autonomous, adaptive, reactive, pro-active, social, cooperative, collaborative and flexible. They also support temporal continuity and mobility within the network. An intelligent agent with mobility feature is known as Mobile Agent (MA). MA migrates from node to node in a heterogeneous network without losing its operability. It can continue to function even if the user is disconnected from the network. On reaching at a node MA is delivered to an Agent Execution Environment (AEE) where its executable parts are started running. Upon completion of the desired task, it delivers the results to the home node. With MA, a single serialized object is transmitted over the network carrying the small amount of resultant data only thus reducing the consumption of network bandwidth, latency (response time delay) and network traffic. An AEE or Mobile Agent Platform (MAP), is server application that provides the appropriate functionality to MAs to authenticate, execute, communicate (with other agents, users, and other platforms), migrate to other platform, and use system resources in a secure way. A Multi Agent System (MAS) is distributed application comprised of multiple interacting intelligent agent components [8].

Table 1.  Single Letter codes for Amino Acids

| Sr. No. | Amino Acid Name | Single Letter Code | Three Letter Code |
|---|---|---|---|
| 1 | Alanine | A | Ala |
| 2 | Cysteine | C | Cys |
| 3 | Aspartic Acid | D | Asp |
| 4 | Glutamic Acid | E | Glu |
| 5 | Phenylalanine | F | Phe |
| 6 | Glycine | G | Gly |
| 7 | Histidine | H | His |
| 8 | Isoleucine | I | Ile |
| 9 | Lysine | K | Lys |
| 10 | Leucine | L | Leu |
| 11 | Methionine | M | Met |
| 12 | Asparagine | N | Asn |
| 13 | Proline | P | Pro |
| 14 | Glutamine | Q | Gln |
| 15 | Arginine | R | Arg |
| 16 | Serine | S | Ser |
| 17 | Threonine | T | Thr |
| 18 | Valine | V | Val |
| 19 | Tryptophan | W | Trp |
| 20 | Tyrosine | Y | Tyr |

Bioinformatics or computational molecular biology aims at automated analysis and the management of high-throughput biological data as well as modelling and simulation of complex biological systems. Bioinformatics has very much changed since the first sequence alignment algorithms in 1970s [9]. Today in-silico analysis is a fundamental component of biomedical research. Bioinformatics has now encompasses a wide range of subject areas from structural biology, genomics to gene expression studies [10], [11]. The post-genomic era has resulted into availability of the enormous amount of distributed biological data sets that require suitable tools and methods for modelling and analyzing biological processes and sequences. The bioinformatics research community feels a strong need to develop new models and exploit and analyze the available genomes [12]. The protein sequences are made up of 20 types of Amino Acids (AA). Each AA is represented by a single letter alphabet as shown in Table 1. Unique 3-dimensional structure of each protein is decided completely by its amino-acid sequence. Proteins are important constituents of cellular machinery of any organism and the functioning of proteins heavily depends upon its AA sequence. A slight change in the sequence might completely change the functioning of the protein. Therefore, the nature of associations between different amino acids has been a subject of great anxiety. In [20], authors mine the association rules among amino acids in protein sequences. As per the literature available, no researcher has applied the agent technology to mine association rules for amino acids from distributed protein data sets. The amalgamation of the DDM and MAS provides rewarding solution in terms of security, scalability, storage cost, computation cost and communication cost. Mining biological data is an emerging area and continues to be an extremely important problem, both for DM and for biological sciences [21]. State of the art in use of agent technology in bioinformatics is reviewed in [22].

The rest of the paper is organised as follows. Section 2 discusses the distributed association rule mining and preliminary notations used in this paper. A running environment for the proposed system is present in Section 3 along with algorithms used for various components and protein data banks used in this study. The layout and working of various agents involved in the proposed system and their algorithms are also discussed. System is implemented in Java and its performance is studied in Section 4 and finally the article is concluded in Section 5.

## 2. DISTRIBUTED ASSOCIATION RULE MINING

Let $DB = \{T_j, j = 1 \ldots D\}$ be a transactional dataset of size $D$ where each transaction $T$ is assigned an identifier ($TID$) and $I = \{d_i, i = 1 \ldots m\}$, total $m$ data items in $DB$. A set of items in a particular transaction $T$ is called itemset or pattern. An itemset, $P = \{d_i, i = 1 \ldots k\}$, which is a set of $k$ data items in a particular transaction $T$ and $P \subseteq I$, is called k-itemset. Support of an itemset, $s(P) = \dfrac{\text{No\_of\_T\_containing\_P}}{D} \%$ is the frequency of occurrence of itemset $P$ in $DB$, where No\_of\_T\_containing\_P is the support count (sup\_count) of itemset $P$. Frequent Itemsets (FIs) are the itemset that appear in $DB$ frequently, i.e., if $s(P) \geq \min\_th\_sup$ (given minimum threshold support), then $P$ is a frequent k-itemset. Finding such FIs plays an essential role in miming the interesting relationships among itemsets. Frequent Itemset Mining (FIM) is the task of finding the set of all the subsets of FIs in a transactional database. It is CPU and input/output intensive task, mainly because of the large size of the datasets involved [2].

Association Rules (ARs) first introduced in [13], are used to discover the associations (or co-occurrences) among item in a database. AR is an implication of the form $P \Rightarrow Q \left[\text{support,confidence}\right]$ where, $P \subset I, Q \subset I$ and $P$ and $Q$ are disjoint itemsets, i.e., $P \cap Q = \varnothing$. An AR is measured in terms of its support ($s$) and confidence ($c$) factors. An AR $P \Rightarrow Q$ is said to be strong if $s(P \Rightarrow Q) \geq \min\_th\_sup$ (given minimum threshold support) and $c(P) \geq \min\_th\_conf$ (given minimum threshold confidence). Association Rule Mining (ARM) today is one of the most important aspects of DM tasks. In ARM all the strong ARs are generated from the FIs. The ARM can be viewed as two step process [14], [15].

1. Find all the frequent k-itemsets($L_k$)
2. Generate Strong ARs from $L_k$
   a. For each frequent itemset, $l \in L_k$, generate all non empty subsets of $l$.
   b. For every non empty subset $s$ of $l$, output the rule "$s \Rightarrow (l - s)$", if $\dfrac{\text{sup\_count}(l)}{\text{sup\_count}(s)} \geq \min\_th\_conf$

Distributed Association Rule Mining (DARM) generates the globally strong association rules from the global FIs in a distributed environment. Because of an intrinsic data skew property of the distributed database, it is desirable to mine the global rules for the global business decisions and the local rules for the local business decisions. Comparative analysis of existing agent based DARM systems can be found in [23].

Few preliminaries notations and definitions required for defining DARM and to make this study self contained are as follows:

- $S = \{S_i, i = 1\ldots n\}$, $n$ distributed sites.

- $S_{CENTRAL}$, Central Site.

- $PDB_i = \{PR_m, m = 1\ldots X_i\}$, Protein Data Bank of $X_i$ protein records at site $S_i$, each $PR_m$ has two main parts; first part is the description headers ($PD$) for structural classification of protein and the second part contains protein sequence ($PS$), i.e., the chain of amino acids. Snapshot of $PDB_I$ is shown in Appendix A.1.

- $FPDB_i = \{PR_k, k = 1\ldots D_i\}$, Filtered Protein Data Bank of $D_i$ protein records at site $S_i$, where length of $PS$ in each $PR$ is in the range $\geq 50$ and $< 400$. Snapshot of $FPDB_I$ is shown in Appendix A.2.

- $AAF_i$, Data bank of amino acids frequency for each $PR \in FPDB$ at site $S_i$. $AAF_1$ is shown in Appendix A.3. $AAF_2$ and $AAF_3$ are not shown due to space constraint.

- $BDB_i$, Boolean Data Bank that contains a value '1' if the frequency of an amino acid lies in the specific range and '0' otherwise at site $S_i$. Snapshot of $BDB_1$ is shown in Appendix A.5.

- $IDB_i$, Itemset Data Bank of frequency partition items to map boolean value '1' with its frequency partition item number at site $S_i$. Snapshot of $IDB_1$ is shown in Appendix A.6.

- $L_{k(i)}^{FI}$, Local frequent k-itemsets at site $S_i$.

- $L_{k(i)}^{FISC}$, List of support count $\forall Itemset \in L_{k(i)}^{FI}$.

- $L_i^{LSAR}$, List of locally strong association rules at site $S_i$.

- $L^{TLSAR} = \bigcup_{i=1}^{n} L_i^{LSAR}$, List of total locally strong association rules.

- $L_k^{TFI} = \bigcup_{i=1}^{n} L_{k(i)}^{FI}$, List of total frequent k-itemsets.

- $L_k^{GFI} = \bigcap_{i=1}^{n} L_{k(i)}^{FI}$, List of global frequent k-itemsets.

- $L_{CENTRAL}^{GSAR}$, List of Globally strong association rule.

Local Knowledge Base (LKB), at site $S_i$, comprises of $L_{k(i)}^{FI}$, $L_{k(i)}^{FISC}$ and $L_i^{LSAR}$ which can provide reference to the local supervisor for local decisions. Global Knowledge Base (GKB), at $S_{CENTRAL}$, comprises of $L^{TLSAR}$, $L_k^{TFI}$, $L_k^{GFI}$ and $L_{CENTRAL}^{GSAR}$ for the global decision making. Like ARM, DARM task can also be viewed as two-step process [15]:

1. Find the global frequent k-itemset ($L_k^{GFI}$) from the distributed Local frequent k-itemsets ($L_{k(i)}^{FI}$) from the partitioned datasets.

2. Generate globally strong association rules ($L_{CENTRAL}^{GSAR}$) from $L_k^{GFI}$.

# 3. PROPOSED AEQARM-AAPDB SYSTEM

## 3.1. Environment for the proposed system

Every MAS needs an underlying AEE to provide running infrastructure on which agents can be deployed and tested. A running environment has been designed in Java to execute all the DM agents involved in the proposed system. Various attributes of the MA are encapsulated within a data structure known as *AgentProfile*. *AgentProfile* contains the name of MA (*AgentName*), version number (*AgentVersion*), entire byte code (*BC*), list of nodes to be visited by MA, i.e., itinerary plan ($L^{NODES}$), type of the itinerary (*ItinType*) which can be serial or parallel, a reference of current execution state (*AObject*) and an additional data structure known as *Briefcase* that acts as a result bag of MA to store final resultant knowledge (*Result_$S_i$*) at a particular site. Computational time (*CPUTime*) taken by a MA at a particular site is also stored in *Result_$S_i$*. In addition to results, *Briefcase* also contains the system time for start of agent journey (*TripTime$_{start}$*), system time for end of journey (*TripTime$_{end}$*) and total round trip time of MA (*TripTime*) calculated using the formula $TripTime \leftarrow TripTime_{end} - TripTime_{start}$. This environment consists of the following three components:

- **Data Mining Agent Execution Environment (DM_AEE):** It is the key component that acts as a Server. DM_AEE is deployed on any distributed sites $S_i$ and is responsible for receiving, executing and migrating all the visiting data mining (DM) agents. It receives the incoming *AgentProfile* at site $S_i$, retrieves the entire *BC* of agent and save it with *AgentName.class* in the local file system of the site $S_i$ after that execution of the agent is started using *AObject*. Steps are shown in Algorithm 1.

- **Agent Launcher (AL)**: It acts a Client at agent launching station ($S_{CENTRAL}$) and launches the goal oriented DM agents on behalf of the user through a user interface to the DM_AEE running at the distributed sites. Agent Pool (or Zone) at $S_{CENTRAL}$ is a repository of all mobile as well as stationary agents (SAs). AL first reads and stores *AgentName* in *AgentProfile*. The entire *BC* of the *AgentName* is loaded from the Agent Pool and stored in *AgentProfile*. $L^{NODES}$ and *ItinType* are retrieved and stored in *AgentProfile*. *TripTime$_{start}$* is maintained in *Briefcase* which is further added to *AgentProfile*. In case of parallel computing model, i.e., if *ItinType* = *Parallel* AL creates $L^{NODES}$ number of clones of the specific MA and dispatches each clone in parallel to all the sites listed in $L^{NODES}$. Here $L^{NODES}$ number of threads are created to dispatch the MAs in parallel. Each clone has *AgentVersion* starting from 1 to size of $L^{NODES}$, which is used to identify each clone on the network. Before dispatching the clone of a MA to DM_AEE, the current state of the newly created ith clone object (*AObject*[$i$]) is also stored in *AgentProfile*. AL also contacts the Result Manager (RM) for processing the *Briefcase* of an agent. Detailed steps are given in Algorithm 2.

- **Result Manager (RM)**: It manages and processes the *Briefcase* of all MAs. RM is either contacted by a MA for submitting its results or by AL for processing the results of the specific MA. On completion of itinerary, each DM agent submits its results to RM which computes total round trip time (*TripTime*) of that MA and saves it in the *Briefcase* of that agent. If *ItinType* = *Parallel* then it saves the *AgentProfile* of all the clones of the agent with *AgentName* in a collection $L_{AgentName}^{AllProfiles}$. It is assumed that all the clones report their results to RM. RM may be equipped with the feature of non reporting clones by issuing an alert to AL for that clone with specific *AgentVersion* . AL then launch a new clone for the specific *AgentVersion* for the specific site. When it is contacted by AL for processing the results of a specific agent it sends back a collection $L_{AgentName}^{AllProfiles}$ for all the clones of that agent. Steps are defined in Algorithm 3.

---

**Algorithm 1** Data Mining Agent Execution Environment (DM_AEE)

1: **procedure** DM_AEE( )
2: **while** TRUE **do**
3:     *AgentPofile ← listen and receive AgentProfile at $S_i$*
4:     *AgentName ← get AgentName from AgentProfile*
5:     *BC ← retrieve the BC of agent from AgentProfile*
6:     *save the BC with AgentName.class in the local file system of $S_i$*
7:     *AObject ← get AObject from AgentProfile*        ▷ current state
8:     *AObject.run()*        ▷ start executing mobile agent
9:    **end while**
10: **end procedure**

---

**Algorithm 2** Agent Launcher (AL) for AeQARM-AAPDB

1: **procedure** AL( )
2: *option ← read option(dispatch / result)*
3: **switch** *option* **do**
4:    **case** *dispatch*        ▷ dispatch the mobile agent to DM_AEE
5:     *AgentName ← read Mobile Agent's name*
6:     *BC ← load entire byte code of AgentName from AgentPool*
7:     *add AgentName and BC to AgentProfile*
8:     *$L^{NODES}$ ← read Itinerary(IP addresses) of mobile agent*
9:     *ItinType ← read ItinType( Serial / Parallel)*
10:     *add ItinType to AgentProfile*
11:    **if** *ItinType* = "Parallel" **then**        ▷ Parallel Itinerary
12:     *AObject[$L^{NODES}$.size]*        ▷ Array of Agent Objects for clone references
13:     *TripTime$_{start}$ ← get system time for start of the agent journey*
14:     *add TripTime$_{start}$ to Briefcase*
15:     *add Breifcase to AgentProfile*
16:    **switch** *AgentName* **do**
17:     **case** *PDBFA*
18:      **for** *i ← 1, $L^{NODES}$.size* **do**        ▷ for each node in the itinerary
19:       *AgentVersion ← i*
20:       *add AgentVersion to AgentProfile*
21:       *NodeAddress ← $L^{NODES}$.get(i)*        ▷ get an IP address
22:       *AObject[i] ← new PDBFA(AgentProfile)*
23:       *Add AObject[i] to AgentProfile*        ▷ clone's state
24:       *Transfer AgentProfile to DM_AEE at NodeAddress*
25:      **end for**

```
26:          end case
27:          case  AAFFA
28:              for i ← 1, L^NODES.size do                          ▷ for each node in the itinerary
29:                  AgentVersion ← i
30:                  add AgentVersion to AgentProfile
31:                  NodeAddress ← L^NODES.get(i)                           ▷ get an IP address
32:                  AObject[i] ← new AAFFA(AgentProfile)
33:                  Add AObject[i] to AgentProfile                        ▷ clone's state
34:                  Transfer AgentProfile to DM_AEE at NodeAddress
35:              end for
36:          end case
37:          case  FMIDBGA
38:              max_freq ← read maximum frequency range for amino acids
39:              for i ← 1, L^NODES.size do                          ▷ for each node in the itinerary
40:                  AgentVersion ← i
41:                  add AgentVersion to AgentProfile
42:                  NodeAddress ← L^NODES.get(i)                           ▷ get an IP address
43:                  AObject[i] ← new FMIDBGA(AgentProfile, max_freq)
44:                  Add AObject[i] to AgentProfile                        ▷ clone's state
45:                  Transfer AgentProfile to DM_AEE at NodeAddress
46:              end for
47:          end case
48:          case  LKGA_P
49:              min_s ← read minimum threshold support
50:              min_c ← read minimum threshold confidence
51:              for i ← 1, L^NODES.size do                          ▷ for each node in the itinerary
52:                  AgentVersion ← i
53:                  add AgentVersion to AgentProfile
54:                  NodeAddress ← L^NODES.get(i)                           ▷ get an IP address
55:                  AObject[i] ← new LKGA_P(AgentProfile, min_s, min_c)
56:                  Add AObject[i] to AgentProfile                        ▷ clone's state
57:                  Transfer AgentProfile to DM_AEE at NodeAddress
58:              end for
59:          end case
60:          case  LKCA_P
61:              for i ← 1, L^NODES.size do                          ▷ for each node in the itinerary
62:                  AgentVersion ← i
63:                  add AgentVersion to AgentProfile
64:                  NodeAddress ← L^NODES.get(i)                           ▷ get an IP address
65:                  AObject[i] ← new LKCA_P(AgentProfile)
66:                  Add AObject[i] to AgentProfile                        ▷ clone's state
67:                  Transfer AgentProfile to DM_AEE at NodeAddress
68:              end for
69:          end case
70:          case  GKDA_P
71:              L^GSAR_CENTRAL ← load L^CENTRAL_CANTRAL generated by GKGA at S_CENTRAL
```

72:        *add $L_{CENTRAL}^{GSAR}$ to Briefcase*

73:        *add updated Briefcase to AgentProfile*

74:       **for** *i ← 1, $L^{NODES}$.size* **do**            ▷ *for each node in the itinerary*

75:         *AgentVersion ← i*

76:         *add AgentVersion to AgentProfile*

77:         *NodeAddress ← $L^{NODES}$.get(i)*         ▷ *get an IP address*

78:         *AObject[i] ← new GKDA_P(AgentProfile)*

79:         *Add AObject[i] to AgentProfile*         ▷ *clone's state*

80:         *Transfer AgentProfile to DM_AEE at NodeAddress*

81:      **end for**

82:     **end case**

83:    **end switch**

84:   **end if**

85:  **end case**

86: **case** *result*                 ▷ *process the results of mobile agent*

87:   *AgentName ← read mobile agent's name*

88:   *ItinType ← read mobile agent's itinerary type*

89:   *add AgentName and ItinType to $L^{AgentInfo}$*

90:   **if** *ItinType = "Parallel"* **then**

91:    *$L_{AgentName}^{AllProfile}$ ← contact RM for $L^{AgentInfo}$*

92:    **switch** *AgentName* **do**

93:     **case** *PDBFA*

94:      **for all** *AgentProfile ∈ $L_{AgentName}^{AllProfile}$* **do**      ▷ *for each clone*

95:       *Briefcase ← retrieve Briefcase from AgentProfile*

96:       *process the Briefcase of PDBFA clone*

97:      **end for**

98:     **end case**

99:     **case** *AAFFA*

100:      **for all** *AgentProfile ∈ $L_{AgentName}^{AllProfile}$* **do**      ▷ *for each clone*

101:       *Briefcase ← retrieve Briefcase from AgentProfile*

102:       *process the Briefcase of AAFFA clone*

103:      **end for**

104:     **end case**

105:     **case** *FMIDBGA*

106:      **for all** *AgentProfile ∈ $L_{AgentName}^{AllProfile}$* **do**      ▷ *for each clone*

107:       *Briefcase ← retrieve Briefcase from AgentProfile*

108:       *process the Briefcase of FMIDBGA clone*

109:      **end for**

110:     **end case**

111:     **case** *LKGA_P*

112:      **for all** *AgentProfile ∈ $L_{AgentName}^{AllProfile}$* **do**      ▷ *for each clone*

113:       *Briefcase ← retrieve Briefcase from AgentProfile*

114:       *process the Briefcase of LKGA_P clone*

115:      **end for**

116:     **end case**

117:     **case** *LKCA_P*

118:      *call RIGKGA($L_{AgentName}^{AllProfile}$)*         ▷ *stationary agent*

119:           **end case**

120:           **case** $GKDA\_P$

121:               **for all** $AgentProfile \in L_{AgentName}^{AllProfile}$ **do**           ▷ for each clone

122:                    $Briefcase \leftarrow$ *retrieve Briefcase from AgentProfile*

123:                    *process the Briefcase of GKDA_P clone*

124:               **end for**

125:           **end case**

126:         **end switch**

127:       **end if**

128:     **end case**

129:   **end switch**

130: **end procedure**

---

**Algorithm 3** RESULT MANAGER(RM)

1:   **procedure RM( )**

2:   **while** TRUE **do**

3:     *listen and receive the incomming request*

4:     **if** *contacted by a mobile agent for submitting results from site* $S_i$ **then**

5:       $AgentProfile \leftarrow$ *receive the incomming AgentProfile from site* $S_i$

6:       $ItinType \leftarrow$ *retrieve ItinType from AgentProfile*

7:       $Briefcase \leftarrow$ *retrieve mobile agent's Briefcase from AgentProfile*

8:       $TripTime_{start} \leftarrow$ *retrieve* $TripTime_{start}$ *from Briefcase*

9:       $TripTime_{end} \leftarrow$ *retrieve* $TripTime_{end}$ *from Briefcase*

10:      $TripTime \leftarrow TripTime_{end} - TripTime_{start}$

11:      *add TripTime to Briefcase*

12:      *add updated Briefcase to AgentProfile*

13:      **if** $ItinType =$ "Parallel" **then**

14:        $AgentName \leftarrow$ *retrieve AgentName from AgentProfile*

15:        $AgentVersion \leftarrow$ *retrieve AgentVersion from AgentProfile*

16:        **if** $AgentVersion = 1$ **then**

17:          *add AgentProfile to* $L_{AgentName}^{AllProfiles}$

18:          *save* $L_{AgentName}^{AllProfiles}$ *at* $S_{CENTRAL}$

19:        **end if**

20:        **if** $AgentVersion > 1$ **then**

21:          *retrieve* $L_{AgentName}^{AllProfiles}$ *from* $S_{CENTRAL}$

22:          *add AgentProfile to* $L_{AgentName}^{AllProfiles}$

23:          *save updated* $L_{AgentName}^{AllProfiles}$ *at* $S_{CENTRAL}$

24:        **end if**

25:      **end if**

26:     **end if**

27:     **if** *contacted by AgentLauncher for processing the results* **then**

28:       $AgentName \leftarrow$ *retrieve AgentName from AgentProfile*

29:       $ItinType \leftarrow$ *retrieve ItinType from incomming* $L^{AgentInfo}$

30:       **if** $ItinType =$ "Parallel" **then**

31:         $L_{AgentName}^{AllProfiles} \leftarrow load\ L_{AgentName}^{AllProfiles}\ from\ S_{CENTRAL}$

32:      *dispatch* $L_{AgentName}^{AllProfiles}$ *to AgentLauncher*
33:      **end if**
34:      **end if**
35:      **end while**
36: **end procedure**

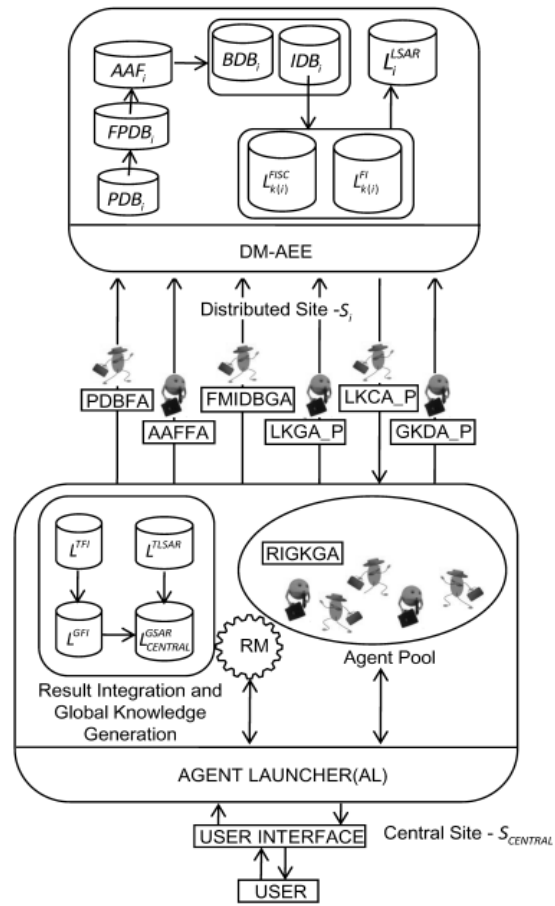

Figure 1. AeQARM-AAPDB MAS

## 3.2. Protein Data Bank

AeQARM-AAPDB system is tested on a real dataset of protein sequences, the Astral SCOP [16], [17 ] version 1.75 genetic domain sequence subsets, based on PDB SEQRES records with less than 40% identity to each other [18]. There are total of 10569 Protein records in this dataset. This single PDB is divided into 3 unites ( $PDB_1$, $PDB_2$ and $PDB_3$) of 3523 protein records in each and are stored at three distributed sites. Each $PDB_i$ is further filtered to generate $FPDB_i$ for $PS$ length range $\geq 50$ and $\leq 400$. A total of only 9633 (3341 ( $FPDB_1$) + 3253 ( $FPDB_2$) + 3039 ( $FPDB_3$)) such filtered records are considered for the mining. The frequencies of 20 amino acids in each protein sequence are retrieved and stored in $AAF_i$ which is further mapped into $BDB_i$ for each amino acid having 15 frequency ranges (partitions) resulting into 300 amino acids $\langle attribute, value \rangle$ pairs as shown in Appendix A.4.

### 3.3. Layout and working of AeQARM-AAPDB system

AeQARM-AAPDB MAS is shown in Figure 1. This MAS consists of total seven agents, clones of six MAs in serial number 1 to 6 are dispatched from $S_{CENTRAL}$ with parallel itinerary migration and one at serial number 7 is an intelligent stationary agent (SA) running at $S_{CENTRAL}$ to perform the different tasks. The CPU time taken by a MA while processing on each site along with some other specific information is carried back in the result bag at $S_{CENTRAL}$. Relationship among these agents and their working behaviour are given as follows.

1. ***Protein Data Bank Filtering Agent (PDBFA)***: Clones of this MA is dispatched in parallel to each distributed site by AL. It carries the *AgentProfile* along with it and filters $PDB_i$ to generate $FPDB_i$ at each site $S_i$. PDBFA carries back the computational time ($CPUTime$) at each site $S_i$ and $TripTime_{end}$.

2. ***Amino Acids Frequency Finder Agent (AAFFA):*** Every clone of this MA carries the *AgentProfile* along with it and finds the frequencies of each amino acids in every protein sequence record in $FPDB_i$ to create $AAF_i$ at each site $S_i$. AAFFA carries back the computational time ($CPUTime$) at each site $S_i$ and $TripTime_{end}$.

3. ***Frequency Mapping and Itemset Data Bank Generater Agent (FMIDBGA)***: Every clone of this MA carries the *AgentProfile* and *maxfrq* (the given maximum frequency range for amino acids) along with it. It divides the frequencies of each amino acids in $AAF_i$ into intervals and maps the frequencies into Boolean values to create $BDB_i$ for frequency intervals and further maps it to $IDB_i$ for frequency interval items.

4. ***Local Knowledge Generator Agent with Parallel Itinerary (LKGA_P)***: Every cloned LKGA_P carries the *AgentProfile*, *min_th_sup* and *min_th_conf* along with it. This agent is embedded with Apriori algorithm [19] for generating all the frequent k-itemset lists. This agent may be equipped with decision making capability to select other FIM algorithms based on the density of the dataset at a particular site. It first performs the FIM to generate and store $L_{k(i)}^{FI}$ and $L_{k(i)}^{FISC}$ at site $S_i$ by scanning the local $IDB_i$ at that site with the constraint of *min_th_sup*. It then performs the ARM applying the constraint of *min_th_conf* to generate and store $L_i^{LSAR}$ by using the $L_{k(i)}^{FI}$ and $L_{k(i)}^{FISC}$. $L_i^{LSAR}$ list also contains support and confidence for a particular association rule along with site name. It carries back the computational time ($CPUTime$) at each site $S_i$ and $TripTime_{end}$.

5. ***Local Knowledge Collector Agent with Parallel Itinerary (LKCA_P)***: Every cloned LKGA_P carries the *AgentProfile* and collects the list of local frequent k-itemset ($L_{k(i)}^{FI}$) and list of locally strong association rules ($L_i^{LSAR}$) generated by LKGA_P. It carries back these distributed results in the result bag to RM at $S_{CENTRAL}$ where these results are integrated with the help of a RIA stationary agent. In addition to this resultant knowledge it also carries back the computational time ($CPUTime$) at each site $S_i$ and $TripTime_{end}$.

6. ***Global Knowledge Dispatcher Agent with Parallel Itinerary (GKDA_P)***: Every cloned GKDA_P carries *AgentProfile* containing global knowledge ($L_{CENTRAL}^{GSAR}$) for further decision making and comparing with local knowledge at that site. It carries back the computational time ($CPUTime$) at each site $S_i$ and $TripTime_{end}$.

7. **_Result Integration and Global Knowledge Generater Agent (RIGKGA)_**: It is a stationary agent at $S_{CENTRAL}$, mainly used for processing the result bags of all clones of LKCA_P. It creates a list of total frequent k-itemset ($L_k^{TFI}$), a list of global frequent itemset, $L_k^{GFI}$ and a list of total locally strong association rules ($L^{TLSAR}$). $L_k^{TFI}$ and $L^{TLSAR}$ are further processed to generate and store $L_{CENTRAL}^{GSAR}$ list.

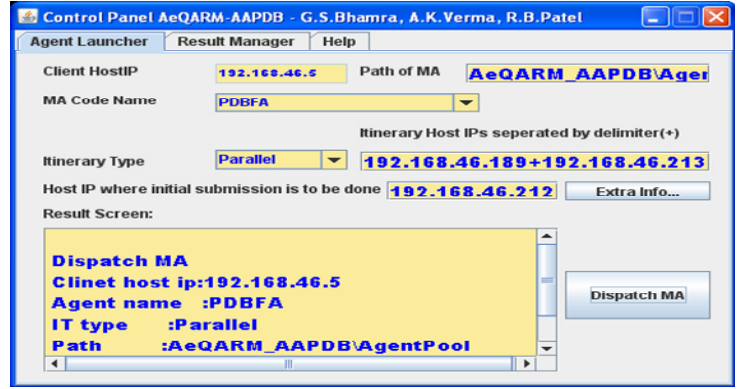

Figure 2.  Control Panel of AeQARM-AAPDB

Table 2.  Network Configuration

| Site Name | Processor | OS | LAN Configuration | |
|---|---|---|---|---|
| | | | IP [a] | Network |
| $S_{CENTRAL}$ | Intel [b] | MS [c] | 192.168.46.5 | NW [d] |
| $S_1$ | Intel [b] | MS [c] | 192.168.46.212 | NW [d] |
| $S_2$ | Intel [b] | MS [c] | 192.168.46.189 | NW [d] |
| $S_3$ | Intel [b] | MS [c] | 192.168.46.213 | NW [d] |
| a. IP address with Mask: 255.255.255.0 and Gateway 192.168.46.1 | | | | |
| b. Intel Pentium Dual Core(3.40 GHz, 3.40 GHz) with 512 MB RAM | | | | |
| c. Microsoft Windows XP Professional ver. 2002 | | | | |
| d. Network Speed: 100 Mbps and Network Adaptor: 82566DM-2 Gigabit NIC | | | | |

## 4. IMPLEMENTATION AND PERFORMANCE STUDY

All the agents as well as Control Panel of the system, as shown in Figure 2, are implemented in Java. The required configuration for the deployment of the system is shown in Table 2 with additional deployment of DM_AEE at each distributed site and AL and RM at $S_{CENTRAL}$.

| Sr. No. | L | AR (support,confidence) | Site |
|---|---|---|---|
| 1 | | [92] => [16] ( 23%, 73% ) | 192.168.46.212 |
| 2 | [16, 92] | [92] => [16] ( 24%, 66% ) | 192.168.46.189 |
| 3 | | [92] => [16] ( 23%, 66% ) | 192.168.46.213 |
| 4 | | [16] => [151] ( 44%, 59% ) | 192.168.46.212 |
| 5 | [16, 151] | [151] => [16] ( 44%, 82% ) | 192.168.46.212 |
| 6 | | [151] => [16] ( 20%, 72% ) | 192.168.46.189 |
| 7 | | [151] => [16] ( 29%, 64% ) | 192.168.46.213 |
| 8 | | [152] => [16] ( 24%, 72% ) | 192.168.46.212 |
| 9 | [16, 152] | [152] => [16] ( 23%, 62% ) | 192.168.46.189 |
| 10 | | [152] => [16] ( 25%, 68% ) | 192.168.46.213 |
| 11 | | [16] => [271] ( 58%, 78% ) | 192.168.46.212 |
| 12 | | [271] => [16] ( 58%, 79% ) | 192.168.46.212 |
| 13 | [16, 271] | [16] => [271] ( 38%, 64% ) | 192.168.46.189 |
| 14 | | [271] => [16] ( 38%, 66% ) | |
| 15 | | [16] => [271] ( 47%, 73% ) | 192.168.46.213 |
| 16 | | [271] => [16] ( 47%, 67% ) | |
| 17 | [16, 287] | [287] => [16] ( 27%, 78% ) | 192.168.46.212 |
| 18 | | [287] => [16] ( 20%, 66% ) | 192.168.46.189 |
| 19 | | [287] => [16] ( 24%, 68% ) | 192.168.46.213 |
| 20 | | [91] => [271] ( 43%, 82% ) | 192.168.46.212 |
| 21 | | [271] => [91] ( 43%, 59% ) | 192.168.46.212 |
| 22 | [91, 271] | [91] => [271] ( 20%, 77% ) | 192.168.46.189 |
| 23 | | [91] => [271] ( 36%, 83% ) | 192.168.46.213 |
| 24 | | [271] => [91] ( 36%, 52% ) | |
| 25 | | [92] => [271] ( 23%, 71% ) | 192.168.46.212 |
| 26 | [92, 271] | [92] => [271] ( 23%, 64% ) | 192.168.46.189 |
| 27 | | [92] => [271] ( 23%, 67% ) | 192.168.46.213 |
| 28 | | [151] => [271] ( 43%, 80% ) | 192.168.46.212 |
| 29 | | [271] => [151] ( 43%, 59% ) | 192.168.46.212 |
| 30 | [151, 271] | [151] => [271] ( 20%, 69% ) | 192.168.46.189 |
| 31 | | [151] => [271] ( 37%, 81% ) | 192.168.46.213 |
| 32 | | [271] => [151] ( 37%, 53% ) | |
| 33 | | [152] => [271] ( 23%, 70% ) | 192.168.46.212 |
| 34 | [152, 271] | [152] => [271] ( 22%, 60% ) | 192.168.46.189 |
| 35 | | [152] => [271] ( 24%, 66% ) | 192.168.46.213 |
| 36 | | [152] => [271] ( 29%, 80% ) | 192.168.46.212 |
| 37 | [197, 271] | [197] => [271] ( 20%, 77% ) | 192.168.46.189 |
| 38 | | [197] => [271] ( 26%, 76% ) | 192.168.46.213 |
| 39 | | [271] => [287] ( 27%, 77% ) | 192.168.46.212 |
| 40 | [271, 287] | [287] => [271] ( 22%, 71% ) | 192.168.46.189 |
| 41 | | [287] => [271] ( 26%, 75% ) | 192.168.46.213 |

Stong Association Rules for frequent 2-Itemsets

Figure 7. Globally strong association rules for globally frequent 2-itemsets

| Sr. No. | L | AR (support,confidence) | Site |
|---|---|---|---|
| 1 | | <H:3..5> => <C:0..2> ( 23%, 73% ) | 192.168.46.212 |
| 2 | { <C:0..2> <H:3..5> } | <H:3..5> => <C:0..2> ( 24%, 66% ) | 192.168.46.189 |
| 3 | | <H:3..5> => <C:0..2> ( 23%, 66% ) | 192.168.46.213 |
| 4 | | <C:0..2> => <M:0..2> ( 44%, 59% ) | 192.168.46.212 |
| 5 | { <C:0..2> <M:0..2> } | <M:0..2> => <C:0..2> ( 44%, 82% ) | 192.168.46.212 |
| 6 | | <M:0..2> => <C:0..2> ( 20%, 72% ) | 192.168.46.189 |
| 7 | | <M:0..2> => <C:0..2> ( 29%, 64% ) | 192.168.46.213 |
| 8 | | <M:3..5> => <C:0..2> ( 24%, 72% ) | 192.168.46.212 |
| 9 | { <C:0..2> <M:3..5> } | <M:3..5> => <C:0..2> ( 23%, 62% ) | 192.168.46.189 |
| 10 | | <M:3..5> => <C:0..2> ( 25%, 68% ) | 192.168.46.213 |
| 11 | | <C:0..2> => <W:0..2> ( 58%, 78% ) | 192.168.46.212 |
| 12 | | <W:0..2> => <C:0..2> ( 58%, 79% ) | 192.168.46.212 |
| 13 | { <C:0..2> <W:0..2> } | <C:0..2> => <W:0..2> ( 38%, 64% ) | 192.168.46.189 |
| 14 | | <W:0..2> => <C:0..2> ( 38%, 66% ) | |
| 15 | | <C:0..2> => <W:0..2> ( 47%, 73% ) | 192.168.46.213 |
| 16 | | <W:0..2> => <C:0..2> ( 47%, 67% ) | |
| 17 | { <C:0..2> <Y:3..5> } | <Y:3..5> => <C:0..2> ( 27%, 78% ) | 192.168.46.212 |
| 18 | | <Y:3..5> => <C:0..2> ( 20%, 66% ) | 192.168.46.189 |
| 19 | | <Y:3..5> => <C:0..2> ( 24%, 68% ) | 192.168.46.213 |
| 20 | | <H:0..2> => <W:0..2> ( 43%, 82% ) | 192.168.46.212 |
| 21 | | <W:0..2> => <H:0..2> ( 43%, 59% ) | 192.168.46.212 |
| 22 | { <H:0..2> <W:0..2> } | <H:0..2> => <W:0..2> ( 20%, 77% ) | 192.168.46.189 |
| 23 | | <H:0..2> => <W:0..2> ( 36%, 83% ) | 192.168.46.213 |
| 24 | | <W:0..2> => <H:0..2> ( 36%, 52% ) | |
| 25 | | <H:3..5> => <W:0..2> ( 22%, 71% ) | 192.168.46.212 |
| 26 | { <H:3..5> <W:0..2> } | <H:3..5> => <W:0..2> ( 23%, 64% ) | 192.168.46.189 |
| 27 | | <H:3..5> => <W:0..2> ( 23%, 67% ) | 192.168.46.213 |
| 28 | | <M:0..2> => <W:0..2> ( 43%, 80% ) | 192.168.46.212 |
| 29 | | <W:0..2> => <M:0..2> ( 43%, 59% ) | 192.168.46.212 |
| 30 | { <M:0..2> <W:0..2> } | <M:0..2> => <W:0..2> ( 20%, 69% ) | 192.168.46.189 |
| 31 | | <M:0..2> => <W:0..2> ( 37%, 81% ) | 192.168.46.213 |
| 32 | | <W:0..2> => <M:0..2> ( 37%, 53% ) | |
| 33 | | <M:3..5> => <W:0..2> ( 23%, 70% ) | 192.168.46.212 |
| 34 | { <M:3..5> <W:0..2> } | <M:3..5> => <W:0..2> ( 22%, 60% ) | 192.168.46.189 |
| 35 | | <M:3..5> => <W:0..2> ( 24%, 66% ) | 192.168.46.213 |
| 36 | | <Q:3..5> => <W:0..2> ( 29%, 80% ) | 192.168.46.212 |
| 37 | { <Q:3..5> <W:0..2> } | <Q:3..5> => <W:0..2> ( 20%, 73% ) | 192.168.46.189 |
| 38 | | <Q:3..5> => <W:0..2> ( 26%, 76% ) | 192.168.46.213 |
| 39 | | <Y:3..5> => <W:0..2> ( 27%, 77% ) | 192.168.46.212 |
| 40 | { <W:0..2> <Y:3..5> } | <Y:3..5> => <W:0..2> ( 22%, 71% ) | 192.168.46.189 |
| 41 | | <Y:3..5> => <W:0..2> ( 26%, 75% ) | 192.168.46.213 |

Stong Quantitative Association Rules for 2-Itemsets(amino acids)

Figure 8. Globally strong association rules for frequent 2- amino acids

$L_{k(1)}^{FI}$ and $L_{k(1)}^{FISC}$ at site $S_1$ generated by LKGA_P agent with 20% *min_th_sup* are shown in Appendix B.1. Locally strong association rules ($L_1^{LSAR}$) generated by LKGA_P for frequent item numbers at site $S_1$ are shown in Appendix B.2 and the same for their corresponding amino acids frequency range are shown in Appendix B.3. Globally strong association rules ($L_{CENTRAL}^{GSAR}$) for the globally frequent itemsets generated by RIGKGA are shown in Figure 7. When these item numbers are mapped with their corresponding amino acids frequency ranges then globally strong quantitative association rules for frequent 2 amino acids are obtained and shown in Figure 8. The results, as shown in Figure 8, reveals that:

- < Cysteine: 0..2 > is strongly associated with < Histidine: 3..5 >, < Methionine:0..2 >, < Methionine: 3..5 >, < Tryptophan: 0..2 >, and < Tyrosine..3::5 >
- < Tryptophan: 0..2 > is strongly associated with < Histidine: 0..2 >, <Histidine: 3..5 >, < Methionine: 0..2 >, < Methionine: 3..5 >, < Glutamine: 3..5 > and < Tyrosine: 3..5 >

## 5. CONCLUSION

Mobile agents strongly qualify for designing distributed applications. DDM, when clubbed with the agent technology, makes a promising alliance that gives favourable results and provides a rewarding solution in managing Big Data with ever increasing size. In this study, a MAS called AeQARM-AAPDB to mine the strong quantitative association rules among amino acids present in primary structure of the proteins from the distributed proteins data sets using intelligent agents is presented. Such globally strong association rules are used in understanding of protein composition and are desirable for synthesis of artificial proteins. Agent-based bio-data mining leaves the technical details of choosing mining algorithms, forming hybrid system, and preparing specific data format to the intelligent system itself because such requirements are unreasonable for most biologists. It alleviates the technical difficulty while enhance the reusability of the mining algorithms and available datasets. By applying multi-agent based distributed bio-data mining, the computing load can be balanced and the computational effort can be achieved in a parallel and scalable manner.

# AUTHORS

**Gurpreet Singh Bhamra** is currently working as Assistant Professor at Department of Computer Science and Engineering, M. M. University, Mullana, Haryana. He received his B.Sc. (Computer Sc.) and MCA from Kurukshetra University, Kurukshetra in 1995 and 1998, respectively. He is pursuing Ph.D. from Department of Computer Science and Engineering, Thapar University, Patiala, Punjab. He is in teaching since 1998. He has published 12 research papers in International/National Journals and International Conferences. He is a Life Member of Computer Society of India. His research interests are in Distributed Computing, Distributed Data Mining, Mobile Agents and Bio-informatics.

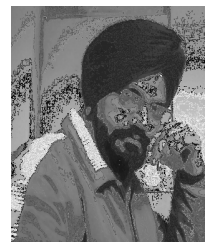

**Dr. Anil Kumar Verma** is currently working as Associate Professor at Department of Computer Science & Engineering, Thapar University, Patiala. He received his B.S., M.S. and Ph.D. in 1991, 2001 and 2008 respectively, majoring in Computer science and engineering. He has worked as Lecturer at M.M.M. Engineering College, Gorakhpur from 1991 to 1996. He joined Thapar Institute of Engineering & Technology in 1996 as a Systems Analyst in the Computer Centre and is presently associated with the same Institute. He has been a visiting faculty to many institutions. He has published over 100 papers in referred journals and conferences (India and Abroad). He is a MISCI (Turkey), LMCSI (Mumbai), GMAIMA (New Delhi). He is a certified software quality auditor by MoCIT, Govt. of India. His research interests include wireless networks, routing algorithms and securing ad hoc networks and data mining.

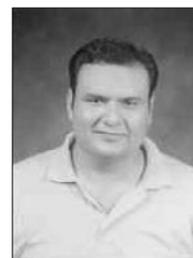

**Dr. Ram Bahadur Patel** is currently working as Professor at Department of Computer Science & Engineering, Chandigarh College of Engineering & Technology, Chandigarh. He received PhD from IIT Roorkee in Computer Science & Engineering, PDF from Highest Institute of Education, Science & Technology (HIEST), Athens, Greece, MS (Software Systems) from BITS Pilani and B. E. in Computer Engineering from M. M. M. Engineering College, Gorakhpur, UP. Dr. Patel is in teaching and research since 1991. He has supervised 36 M. Tech, 7 M. Phil. and 8 PhD Thesis. He is currently supervising 6 PhD students. He has published 130 research papers in International/National Journals and Conferences. He is member of ISTE (New Delhi), IEEE (USA). His current research interests are in Mobile & Distributed Computing, Mobile Agent Security and Fault Tolerance and Sensor Network.

## APPENDIX A – DATASETS USED FOR AeQARM-AAPDB SYSTEM

### A.1 $PDB_1$ at site $S_1$

This is a real dataset of protein sequences, the Astral Structural Classification of Proteins (SCOP) [16], [17] version 1.75 genetic domain sequence subsets, based on PDB SEQRES records with less than 40% identity to each other. SCOP database is comprehensive ordering of all proteins of known structure, according to their evolutionary and structural relationships. Protein domains in the SCOP are grouped into species and hierarchically classified into families, superfamilies, folds and classes. In this dataset each protein record starts with greater than ('>') character followed by protein description headers according to SCOP. An additional hash ('#') character is inserted as a separator character between protein description headers and protein sequence of amino acids. A total of 3523 protein records are stored in this protein data bank ( $PDB_1$ ) at site $S_1$ .

```
>d1d1wa_ a.1.1.1 (A): Protozoan/bacterial hemoglobin (Ciliate (Paramecium caudatum) [TaxId: 5885])#slfeqlg
>d1s69a_ a.1.1.1 (A): Protozoan/bacterial hemoglobin (Cyanobacteria (Synechocystis sp.), pcc 6803 [TaxId:
>d1idra_ a.1.1.1 (A): Protozoan/bacterial hemoglobin (Mycobacterium tuberculosis, HbN [TaxId: 1773])#gllsr
>d1ngka_ a.1.1.1 (A): Protozoan/bacterial hemoglobin (Mycobacterium tuberculosis, HbO [TaxId: 1773])#Ksfyd
>d1ux8a_ a.1.1.1 (A): Protozoan/bacterial hemoglobin (Bacillus subtilis [TaxId: 1423])#napyeaigeellsqlvdtf
>d1kr7a_ a.1.1.4 (A): Nerve tissue mini-hemoglobin (neural globin) (Milky ribbon-worm (Cerebratulus lacteu
>d3sdha_ a.1.1.2 (A): Hemoglobin I (Ark clam (Scapharca inaequivalvis) [TaxId: 6561])#svydaaaqltadvkkdirds
>d1b0ba_ a.1.1.2 (A): Hemoglobin I (Clam (Lucina pectinata) [TaxId: 29163])#slsaaqkdnvksswakasaawgtagpeffm
>d1h97a_ a.1.1.2 (A): Trematode hemoglobin/myoglobin (Paramphistomum epiclitum [TaxId: 54403])#tlthheqdill
>d1jl7a_ a.1.1.2 (A): Glycera globin (Marine bloodworm (Glycera dibranchiata) [TaxId: 6350])#glsaaqrqvvast
>d1a6ma_ a.1.1.2 (A): Myoglobin (Sperm whale (Physeter catodon) [TaxId: 9755])#vlsegewqlvlhvwakveadvaghgqd
>d1mbaa_ a.1.1.2 (A): Myoglobin (Slug sea hare (Aplysia limacina) [TaxId: 6502])#slsaaeadlagkswapvfanknang
>d1ecda_ a.1.1.2 (B:1-146): Erythrocruorin (Midge (Chironomus thummi thummi), fraction III [TaxId: 7154])#lsadqi
>d2qdma_ a.1.1.2 (A): Leghemoglobin (Yellow lupin (Lupinus luteus) [TaxId: 3873])#galtesqaalvkssweefnanipk
>d1cg5a_ a.1.1.2 (A): Hemoglobin, alpha-chain (Cartilaginous fish akaei (Dasyatis akajei) [TaxId: 31902])#
>d1dlka1 a.1.1.2 (A:1-142) Hemoglobin, alpha-chain (Antarctic fish (Trematomus newnesi) [TaxId: 35730])#sl
>d1cg5b_ a.1.1.2 (B): Hemoglobin, beta-chain (Cartilaginous fish akaei (Dasyatis akajei) [TaxId: 31902])#v
>d3d1kb1 a.1.1.2 (B:1-146) Hemoglobin, beta-chain (Antarctic fish (Trematomus newnesi) [TaxId: 35730])#vew
>d1gcvb_ a.1.1.2 (B): Hemoglobin, beta-chain (Houndshark (Mustelus griseus) [TaxId: 89020])#vhwtqeerdeiskt
>d1it2a_ a.1.1.2 (A): Hagfish hemoglobin I)nshore hagfish (Eptatretus burgeri) [TaxId: 7764])#pslidqqglpith
>d1asha_ a.1.1.2 (A): Ascaris hemoglobin, domain I (Pig roundworm (Ascaris suum) [TaxId: 6253])#anktreiceke
>d1itha_ a.1.1.2 (A): Hemoglobin (Innkeeper worm (Urechis caupo) [TaxId: 6431])#gltaaqikaiqdhwflnikgclqaaa
>d1hlba_ a.1.1.2 (A): Hemoglobin, different isoforms (Caudina arenicola, also known as Molpadia arenicola
>d1qlfa_ a.1.1.2 (A): Neuroglobin (Mouse (Mus musculus) [TaxId: 10090])#rpeselsicgswaqvyrxsglehgtvlfarlfrl
>d1cqxa1 a.1.1.2 (A:1-150) Flavohemoglobin, N-terminal domain (Alcaligenes eutrophus [TaxId: 106590])#mitql
>d2qfka1 a.1.1.2 (A:1-137) Dehaloperoxidase (Amphitrite ornata [TaxId: 129555])#gfkqdiatirgdirtyaqdiflafln
>d1or4a_ a.1.1.2 (A): Heme-based aerotactic transducer HemAT, sensor domain (Bacillus subtilis [TaxId: 142
>d1urva_ a.1.1.2 (A): Cytoglobin (Human (Homo sapiens) [TaxId: 9606])#elseaerkavqamswarlyansedvgvailvrffvnf
```

### A.2 $FPDB_1$ at site $S_1$

$PDB_1$ shown in Appendix A.1 is further filtered to generate $FPDB_1$ for the protein sequence length range $\geq 50$ and $< 400$ . A total of 3341 such filtered protein records are obtained.

```
>d1dlwa_ a.1.1.1 (A:) Protozoan/bacterial hemoglobin (Ciliate (Paramecium caudatum) [TaxId: 5885])#slfeqlg
>d1s69a_ a.1.1.1 (A:) Protozoan/bacterial hemoglobin (Cyanobacteria (Synechocystis sp.), pcc 6803 [TaxId:
>d1idra_ a.1.1.1 (A:) Protozoan/bacterial hemoglobin (Mycobacterium tuberculosis, HbN [TaxId: 1773])#gilsr
>d1ngka_ a.1.1.1 (A:) Protozoan/bacterial hemoglobin (Mycobacterium tuberculosis, HbO [TaxId: 1773])#Ksfyd
>d1ux8a_ a.1.1.1 (A:) Protozoan/bacterial hemoglobin (Bacillus subtilis [TaxId: 1423])#napyeaigellsqlvdtf
>d1kr7a_ a.1.1.4 (A:) Nerve tissue mini-hemoglobin (neural globin) (Milky ribbon-worm (Cerebratulus lacteu
>d3sdha_ a.1.1.2 (A:) Hemoglobin I (Ark clam (Scapharca inaequivalvis) [TaxId: 6561])#svydaaaqltadvkkdirds
>d1b0ba_ a.1.1.2 (A:) Hemoglobin I (Clam (Lucina pectinata) [TaxId: 29163])#slaaqkdsvksswkasaawgtagpeffma
>d1h97a_ a.1.1.2 (A:) Trematode hemoglobin/myoglobin (Paramphistomum epiclitum [TaxId: 54403])#tltkheqdill
>d1jl7a_ a.1.1.2 (A:) Glycera globin (Marine bloodworm (Glycera dibranchiata) [TaxId: 6350])#glsaaqrqvvant
>d1a6ma_ a.1.1.2 (A:) Myoglobin (Sperm whale (Physeter catodon) [TaxId: 9755])#vlsegewqlvlhvwakveadvaghgqd
>d1mbaa_ a.1.1.2 (A:) Myoglobin (Slug sea hare (Aplysia limacina) [TaxId: 6502])#slsaaeadlagkswapvfanknang
>d1ecda_ a.1.1.2 (A:) Erythrocruorin (Midge (Chironomus thummi thummi), fraction III [TaxId: 7154])#lsadqi
>d2gdma_ a.1.1.2 (A:) Leghemoglobin (Yellow lupin (Lupinus luteus) [TaxId: 3873])#galtesqaalvkssweefnanipk ●●●
>d1cg5a_ a.1.1.2 (A:) Hemoglobin, alpha-chain (Cartilaginous fish akaei (Dasyatis akajei) [TaxId: 31902])#v
>d3d1ka1 a.1.1.2 (A:1-142) Hemoglobin, alpha-chain (Antarctic fish (Trematomus newnesi) [TaxId: 35730])#sl
>d1cg5b_ a.1.1.2 (B:) Hemoglobin, beta-chain (Cartilaginous fish akaei (Dasyatis akajei) [TaxId: 31902])#vl
>d3d1kb1 a.1.1.2 (B:1-146) Hemoglobin, beta-chain (Antarctic fish (Trematomus newnesi) [TaxId: 35730])#vew
>d1gcvb_ a.1.1.2 (B:) Hemoglobin, beta-chain (Houndshark (Mustelus griseus) [TaxId: 89020])#vhwtqeerdeiskt
>d1it2a_ a.1.1.2 (A:) Hagfish hemoglobin (Inshore hagfish (Eptatretus burgeri) [TaxId: 7764])#piidqqglptlt
>d1asha_ a.1.1.2 (A:) Ascaris hemoglobin, domain 1 (Pig roundworm (Ascaris suum) [TaxId: 6253])#anktrelcmk
>d1itha_ a.1.1.2 (A:) Hemoglobin (Innkeeper worm (Urechis caupo) [TaxId: 6431])#gltaaqikaiqdhwflnikgclqaaa
>d1hlba_ a.1.1.2 (A:) Hemoglobin, different isoforms (Caudina arenicola, also known as Molpadia arenicola
>d1qlfa_ a.1.1.2 (A:) Neuroglobin (Mouse (Mus musculus) [TaxId: 10090])#rpeselirqswvvsrspiehgtvlfarlfalep
>d1cqxa1 a.1.1.2 (A:1-150) Flavohemoglobin, N-terminal domain (Alcaligenes eutrophus [TaxId: 106590])#sltq
>d2qfaa1 a.1.1.2 (A:1-137) Dehaloperoxidase (Amphitrite ornata [TaxId: 129555])#gfkqdiatirgditrqaqdlfalnl
>d1or4a_ a.1.1.2 (A:) Heme-based aerotactic transducer HemAT, sensor domain (Bacillus subtilis [TaxId: 142
>d1urva_ a.1.1.2 (A:) Cytoglobin (Human (Homo sapiens) [TaxId: 9606])#elseaerkavqamwarlyansedvgvailvrffvnf
                                              ⋮
```

## A.3 $AAF_1$ generated from $FPDB_1$ at site $S_1$

Each column heading represents single letter code of an amino acid. Each row represents the frequencies of amino acids in a particular protein sequence record.

```
------------------------------------------------------------------
S. N.| a  c  d  e  f  g  h  i  k  l  m  n  p  q  r  s  t  v  w  y
------------------------------------------------------------------
  1) 22  1  3  4  6 11  2  3  2  8  2  8  2  8  3  3 12 14  1  1
  2) 17  0 12 10  6  9  6  2  9 12  3  3  2  4  6  1  6 10  0  5
  3) 14  0  6  8  6 13  5  8  5 12  3  1  6  4  6  8  7 12  0  3
  4) 13  1 10 12  6  6  5  4  2 12  4  1  5  3 15  8  4  7  2  6
  5)  9  1  5 12  6  7  6  4 16  4  3  9  5  8  4  7  4  1  4
  6) 21  2  7  3  5 12  6  4  9  8  1  6  2  3  6  2  6  1  4
  7) 18  1 12  3  6 10  2  8 15 13  3 10  2  6  4  7  6 13  2  4
  8) 28  1  8  5 10 16  2  2 11  9  6  3  6  3  5  2 11  5  7  4  1
  9) 14  0  6 14  7  8 11 10 14 11  5  3  5  5  3  6 11  9  0  5
 10) 27  1  6  8  4 21  5  7 11 10  5  3  3  4  3 12  1 11  2  3
 11) 17  0  7 14  6 10 12  9 19 18  2  1  4  4  4  6  5  8  2  3
 12) 29  0  8  5 15 11  1  4 11 11  3  9  6  2  4 13  2 10  2  0
 13) 17  0  9  5 14 11  4  9 10  7  5  4  3  9  9  9  1  2
 14) 21  0  6 14  7  7  5 14 14  1  6  5  4  1  9  8 17  3  2
 15) 17  4  7 10  8  5 10  3 10 20  1  9  3  4  4  6  5 10  1  3
 16) 13  1  9  5  5  8 10 14 14  4  4  7  2  5 14  5 11  2  4
 17) 12  0  8  9  9  7  7  6 15 14  1  5  2 10  5  6  6 12  2  5
 18) 15  2 12  5  7 10  7  9  9 13  5  7  3  4  4 10  7  9  2  6
 19)  7  2 15  5  9  5  8  9 12  9  3  1  3  7  6  6 10 12  3  4
 20)  6  2  8 10 10  4  2 17 18 13  1  7  7  8  2 13  5  7  2  4
                                              ⋮
```

## A.4 Amino acids frequency ranges (15 ranges for each amino acid)

Item No. column represents Serial No. from 1 to 300. Data in the AA Frequency Range column represents single letter code of amino acid along with a frequency range. Frequency of each amino acid is divided into 15 partitions (ranges) resulting into 300 items for 20 amino acids. For example if frequency of amino acid Alanine (A) in a protein record is 22 it lies at Item No. 8.

| Item No. | AA Frequency Range | Item No. | AA Frequency Range | Item No. | AA Frequency Range | Item No. | AA Frequency Range | Item No. | AA Frequency Range |
|---|---|---|---|---|---|---|---|---|---|
| 1 | <A:0..2> | 61 | <F:0..2> | 121 | <K:0..2> | 181 | <P:0..2> | 241 | <T:0..2> |
| 2 | <A:3..5> | 62 | <F:3..5> | 122 | <K:3..5> | 182 | <P:3..5> | 242 | <T:3..5> |
| 3 | <A:6..8> | 63 | <F:6..8> | 123 | <K:6..8> | 183 | <P:6..8> | 243 | <T:6..8> |
| 4 | <A:9..11> | 64 | <F:9..11> | 124 | <K:9..11> | 184 | <P:9..11> | 244 | <T:9..11> |
| 5 | <A:12..14> | 65 | <F:12..14> | 125 | <K:12..14> | 185 | <P:12..14> | 245 | <T:12..14> |
| 6 | <A:15..17> | 66 | <F:15..17> | 126 | <K:15..17> | 186 | <P:15..17> | 246 | <T:15..17> |
| 7 | <A:18..20> | 67 | <F:18..20> | 127 | <K:18..20> | 187 | <P:18..20> | 247 | <T:18..20> |
| 8 | <A:21..30> | 68 | <F:21..30> | 128 | <K:21..30> | 188 | <P:21..30> | 248 | <T:21..30> |
| 9 | <A:31..40> | 69 | <F:31..40> | 129 | <K:31..40> | 189 | <P:31..40> | 249 | <T:31..40> |
| 10 | <A:41..50> | 70 | <F:41..50> | 130 | <K:41..50> | 190 | <P:41..50> | 250 | <T:41..50> |
| 11 | <A:51..60> | 71 | <F:51..60> | 131 | <K:51..60> | 191 | <P:51..60> | 251 | <T:51..60> |
| 12 | <A:61..70> | 72 | <F:61..70> | 132 | <K:61..70> | 192 | <P:61..70> | 252 | <T:61..70> |
| 13 | <A:71..80> | 73 | <F:71..80> | 133 | <K:71..80> | 193 | <P:71..80> | 253 | <T:71..80> |
| 14 | <A:81..90> | 74 | <F:81..90> | 134 | <K:81..90> | 194 | <P:81..90> | 254 | <T:81..90> |
| 15 | <A:91..400> | 75 | <F:91..400> | 135 | <K:91..400> | 195 | <P:91..400> | 255 | <T:91..400> |

$\vdots$

## A.5 $BDB_1$ at site $S_1$

This Boolean data bank ($BDB_1$) is created using $AAF_1$ as shown in Appendix A.3 and amino acids frequency range table as shown in Appendix A.4. Each column heading represents an Item No. as shown in Appendix A.4. Each amino acid has 15 frequency partitions and for each amino acid, a boolean value '1' is put under that Item No. in which frequency of amino acid lies in a particular protein record otherwise a value '0' is considered. For example it clear from $AAF_1$ that the frequency of amino acid Alanine (A) in the 1$^{st}$ protein record is 22 and this frequency lies at Item No. 8 in frequency range table as shown in Appendix A.4, so a boolean value '1' is put under Item No. 8 in the 1$^{st}$ protein record in $BDB_1$.

| S.No. | 1 | 2 | 3 | 4 | 5 | 6 | 7 | 8 | 9 | 10 | 11 | 12 | 13 | 14 | 15 | 16 | 17 | 18 | 19 | 20 | 21 | 22 | 23 | 24 | 25 |
|---|---|---|---|---|---|---|---|---|---|---|---|---|---|---|---|---|---|---|---|---|---|---|---|---|---|
| 1) | 0 | 0 | 0 | 0 | 0 | 0 | 0 | 1 | 0 | 0 | 0 | 0 | 0 | 0 | 0 | 1 | 0 | 0 | 0 | 0 | 0 | 0 | 0 | 0 | 0 |
| 2) | 0 | 0 | 0 | 0 | 0 | 0 | 1 | 0 | 0 | 0 | 0 | 0 | 0 | 0 | 0 | 1 | 0 | 0 | 0 | 0 | 0 | 0 | 0 | 0 | 0 |
| 3) | 0 | 0 | 0 | 0 | 1 | 0 | 0 | 0 | 0 | 0 | 0 | 0 | 0 | 0 | 0 | 1 | 0 | 0 | 0 | 0 | 0 | 0 | 0 | 0 | 0 |
| 4) | 0 | 0 | 0 | 0 | 1 | 0 | 0 | 0 | 0 | 0 | 0 | 0 | 0 | 0 | 0 | 1 | 0 | 0 | 0 | 0 | 0 | 0 | 0 | 0 | 0 |
| 5) | 0 | 0 | 0 | 1 | 1 | 0 | 0 | 0 | 0 | 0 | 0 | 0 | 0 | 0 | 0 | 1 | 0 | 0 | 0 | 0 | 0 | 0 | 0 | 0 | 0 |
| 6) | 0 | 0 | 0 | 0 | 0 | 0 | 0 | 1 | 0 | 0 | 0 | 0 | 0 | 0 | 0 | 1 | 0 | 0 | 0 | 0 | 0 | 0 | 0 | 0 | 0 |
| 7) | 0 | 0 | 0 | 0 | 0 | 0 | 1 | 0 | 0 | 0 | 0 | 0 | 0 | 0 | 0 | 1 | 0 | 0 | 0 | 0 | 0 | 0 | 0 | 0 | 0 |
| 8) | 0 | 0 | 0 | 0 | 0 | 0 | 0 | 1 | 0 | 0 | 0 | 0 | 0 | 0 | 0 | 1 | 0 | 0 | 0 | 0 | 0 | 0 | 0 | 0 | 0 |
| 9) | 0 | 0 | 0 | 0 | 1 | 0 | 0 | 0 | 0 | 0 | 0 | 0 | 0 | 0 | 0 | 1 | 0 | 0 | 0 | 0 | 0 | 0 | 0 | 0 | 0 |
| 10) | 0 | 0 | 0 | 0 | 0 | 0 | 0 | 1 | 0 | 0 | 0 | 0 | 0 | 0 | 0 | 1 | 0 | 0 | 0 | 0 | 0 | 0 | 0 | 0 | 0 |
| 11) | 0 | 0 | 0 | 0 | 0 | 0 | 1 | 0 | 0 | 0 | 0 | 0 | 0 | 0 | 0 | 1 | 0 | 0 | 0 | 0 | 0 | 0 | 0 | 0 | 0 |
| 12) | 0 | 0 | 0 | 0 | 0 | 0 | 1 | 0 | 0 | 0 | 0 | 0 | 0 | 0 | 0 | 1 | 0 | 0 | 0 | 0 | 0 | 0 | 0 | 0 | 0 |
| 13) | 0 | 0 | 0 | 0 | 0 | 1 | 0 | 0 | 0 | 0 | 0 | 0 | 0 | 0 | 0 | 1 | 0 | 0 | 0 | 0 | 0 | 0 | 0 | 0 | 0 |
| 14) | 0 | 0 | 0 | 0 | 0 | 0 | 0 | 1 | 0 | 0 | 0 | 0 | 0 | 0 | 0 | 1 | 0 | 0 | 0 | 0 | 0 | 0 | 0 | 0 | 0 |
| 15) | 0 | 0 | 0 | 0 | 0 | 0 | 1 | 0 | 0 | 0 | 0 | 0 | 0 | 0 | 0 | 0 | 1 | 0 | 0 | 0 | 0 | 0 | 0 | 0 | 0 |
| 16) | 0 | 0 | 0 | 1 | 0 | 0 | 0 | 0 | 0 | 0 | 0 | 0 | 0 | 0 | 0 | 1 | 0 | 0 | 0 | 0 | 0 | 0 | 0 | 0 | 0 |
| 17) | 0 | 0 | 0 | 0 | 0 | 0 | 1 | 0 | 0 | 0 | 0 | 0 | 0 | 0 | 0 | 1 | 0 | 0 | 0 | 0 | 0 | 0 | 0 | 0 | 0 |
| 18) | 0 | 0 | 0 | 0 | 0 | 0 | 1 | 0 | 0 | 0 | 0 | 0 | 0 | 0 | 0 | 1 | 0 | 0 | 0 | 0 | 0 | 0 | 0 | 0 | 0 |
| 19) | 0 | 0 | 1 | 0 | 0 | 0 | 0 | 0 | 0 | 0 | 0 | 0 | 0 | 0 | 0 | 1 | 0 | 0 | 0 | 0 | 0 | 0 | 0 | 0 | 0 |
| 20) | 0 | 0 | 1 | 0 | 0 | 0 | 0 | 0 | 0 | 0 | 0 | 0 | 0 | 0 | 0 | 1 | 0 | 0 | 0 | 0 | 0 | 0 | 0 | 0 | 0 |
| 21) | 0 | 0 | 0 | 1 | 0 | 0 | 0 | 0 | 0 | 0 | 0 | 0 | 0 | 0 | 0 | 0 | 1 | 0 | 0 | 0 | 0 | 0 | 0 | 0 | 0 |
| 22) | 0 | 0 | 0 | 0 | 0 | 0 | 0 | 1 | 0 | 0 | 0 | 0 | 0 | 0 | 0 | 1 | 0 | 0 | 0 | 0 | 0 | 0 | 0 | 0 | 0 |
| 23) | 0 | 0 | 0 | 0 | 0 | 0 | 1 | 0 | 0 | 0 | 0 | 0 | 0 | 0 | 0 | 1 | 0 | 0 | 0 | 0 | 0 | 0 | 0 | 0 | 0 |
| 24) | 0 | 0 | 0 | 1 | 0 | 0 | 0 | 0 | 0 | 0 | 0 | 0 | 0 | 0 | 0 | 1 | 0 | 0 | 0 | 0 | 0 | 0 | 0 | 0 | 0 |
| 25) | 0 | 0 | 0 | 0 | 0 | 0 | 0 | 1 | 0 | 0 | 0 | 0 | 0 | 0 | 0 | 1 | 0 | 0 | 0 | 0 | 0 | 0 | 0 | 0 | 0 |
| 26) | 0 | 0 | 0 | 0 | 1 | 0 | 0 | 0 | 0 | 0 | 0 | 0 | 0 | 0 | 0 | 1 | 0 | 0 | 0 | 0 | 0 | 0 | 0 | 0 | 0 |
| 27) | 0 | 0 | 0 | 1 | 0 | 0 | 0 | 0 | 0 | 0 | 0 | 0 | 0 | 0 | 0 | 1 | 0 | 0 | 0 | 0 | 0 | 0 | 0 | 0 | 0 |
| 28) | 0 | 0 | 0 | 0 | 0 | 0 | 0 | 1 | 0 | 0 | 0 | 0 | 0 | 0 | 0 | 1 | 0 | 0 | 0 | 0 | 0 | 0 | 0 | 0 | 0 |
| 29) | 0 | 0 | 0 | 0 | 0 | 0 | 1 | 0 | 0 | 0 | 0 | 0 | 0 | 0 | 0 | 1 | 0 | 0 | 0 | 0 | 0 | 0 | 0 | 0 | 0 |
| 30) | 0 | 0 | 0 | 1 | 0 | 0 | 0 | 0 | 0 | 0 | 0 | 0 | 0 | 0 | 0 | 1 | 0 | 0 | 0 | 0 | 0 | 0 | 0 | 0 | 0 |

$\vdots$

## A.6 $IDB_1$ at site $S_1$

This Itemset data bank ( $IDB_1$ ) is created using $BDB_1$ as shown in Appendix A.4. Each record in this dataset consist of 20 Item Numbers, one for each of amino acids for which a boolean value '1' is stored in $BDB_1$ .

```
 1)    8  16  32  47  63  79  91 107 121 138 151 168 181 198 212 227 245 260 271 286
 2)    6  16  35  49  63  79  93 106 124 140 152 167 181 197 213 226 243 259 271 287
 3)    5  16  33  48  63  80  92 108 122 140 152 166 183 197 213 228 243 260 271 287
 4)    5  16  34  50  63  78  92 107 121 140 152 166 182 197 216 228 242 258 271 288
 5)    4  16  32  50  63  78  93 107 122 141 152 167 184 197 213 227 243 257 271 287
 6)    8  16  33  47  62  80  93 107 124 138 151 168 181 197 211 228 241 258 271 287
 7)    7  16  35  47  63  79  91 108 126 140 152 169 181 198 212 228 243 260 271 287
 8)    8  16  33  47  64  81  91 106 124 139 153 168 182 197 213 229 242 258 272 286
 9)    5  16  33  50  63  78  94 109 125 139 152 167 182 197 212 228 244 259 271 287
10)    8  16  33  48  62  83  92 108 124 139 152 167 182 197 212 230 241 259 271 287
11)    6  16  33  50  63  79  95 109 127 142 151 166 182 197 212 228 242 258 271 287
12)    8  16  33  47  66  79  91 107 124 139 152 169 183 196 212 230 241 259 271 286
13)    6  16  34  47  65  79  92 109 124 138 152 167 182 197 212 229 244 259 271 286
14)    8  16  33  50  63  78  92 109 125 140 151 168 182 197 211 229 243 261 272 286
15)    6  17  33  49  63  77  94 107 124 142 151 169 182 197 212 228 243 259 271 287
16)    6  16  34  47  62  78  92 109 125 140 151 167 182 197 212 229 243 259 271 287
17)    5  16  33  49  64  78  93 108 126 140 151 167 181 199 212 228 243 260 271 287
18)    6  16  35  47  63  79  93 109 124 140 152 168 182 197 212 229 243 259 271 288
19)    3  16  36  47  64  77  93 109 125 139 152 166 182 198 213 228 244 260 272 287
20)    3  16  33  49  64  77  91 111 127 140 151 168 183 198 211 230 242 258 271 287
```

# APPENDIX B – RESULTANT KNOWLEDGE OF AeQARM-AAPDB SYSTEM

## B.1 $L_{k(1)}^{FI}$ and $L_{k(1)}^{FISC}$ at site $S_1$

List of frequent k-itemset, i.e., $L_{k(1)}^{FI}$ is represented by column **L** and column **SC** shows the support count of the corresponding frequent k-itemset, i.e., $L_{k(1)}^{FISC}$ at site $S_1$ . These frequent itemsets and their support counts are obtained by processing the Itemset Data Bank ( $IDB_1$ ) as shown in Appendix A.6.

| Sr. No. | 1-Itemsets | | 2-Itemsets | | 3-Itemsets | | 4-Itemsets | |
|---|---|---|---|---|---|---|---|---|
| | L | SC | L | SC | L | SC | L | SC |
| 1 | [2] | 713 | [16, 32] | 852 | [16, 32, 271] | 735 | [16, 91, 151, 271] | 869 |
| 2 | [3] | 765 | [16, 33] | 731 | [16, 61, 271] | 736 | | |
| 3 | [16] | 2508 | [16, 48] | 769 | [16, 62, 271] | 757 | | |
| 4 | [32] | 1015 | [16, 61] | 838 | [16, 91, 151] | 998 | | |
| 5 | [33] | 923 | [16, 62] | 933 | [16, 91, 271] | 1252 | | |
| 6 | [48] | 955 | [16, 91] | 1473 | [16, 91, 286] | 670 | | |
| 7 | [49] | 695 | [16, 92] | 784 | [16, 107, 271] | 691 | | |
| 8 | [61] | 948 | [16, 107] | 831 | [16, 151, 271] | 1239 | | |
| 9 | [62] | 1172 | [16, 108] | 684 | [16, 167, 271] | 778 | | |
| 10 | [77] | 796 | [16, 123] | 693 | [16, 182, 271] | 769 | | |
| 11 | [78] | 849 | [16, 151] | 1492 | [16, 197, 271] | 800 | | |
| 12 | [91] | 1769 | [16, 152] | 811 | [16, 212, 271] | 728 | | |
| 13 | [92] | 1073 | [16, 166] | 690 | [16, 227, 271] | 680 | | |
| 14 | [107] | 1010 | [16, 167] | 918 | [16, 242, 271] | 724 | | |
| 15 | [108] | 870 | [16, 181] | 674 | [16, 271, 286] | 826 | | |
| 16 | [122] | 767 | [16, 182] | 917 | [16, 271, 287] | 750 | | |
| 17 | [123] | 874 | [16, 197] | 980 | [91, 151, 271] | 999 | | |
| 18 | [138] | 708 | [16, 212] | 859 | [91, 271, 286] | 672 | | |
| 19 | [139] | 682 | [16, 213] | 679 | | | | |
| 20 | [151] | 1813 | [16, 227] | 766 | | | | |
| 21 | [152] | 1112 | [16, 242] | 836 | | | | |
| 22 | [166] | 781 | [16, 271] | 1961 | | | | |
| 23 | [167] | 1145 | [16, 286] | 952 | | | | |
| 24 | [168] | 747 | [16, 287] | 932 | | | | |
| 25 | [181] | 733 | [32, 91] | 670 | | | | |

## B.2 $L_1^{LSAR}$ for Item numbers at site $S_1$

Column **L** represents frequent k-itemset and column **AR(support,confidence)** shows the list of locally strong association rules, i.e., $L_1^{LSAR}$ at site $S_1$. Each strong rule has its associated support and confidence factor. The minimum threshold support is taken as 20% and minimum threshold confidence as 50% for generating the strong rules by making use of the data as shown in Appendix B.1.

| Sr. No. | Strong ARs for Frequent 4-itemsets | | Strong ARs for Frequent 3-itemsets | | Strong ARs for Frequent 2-itemsets | |
|---|---|---|---|---|---|---|
| | L | AR(support,confidence) | L | AR(support,confidence) | L | AR(support,confidence) |
| 1 | | [16, 91] => [151, 271] ( 26%, 58% ) | | [32] => [16, 271] ( 21%, 72% ) | [16, 32] | [32] => [16] ( 25%, 83% ) |
| 2 | | [16, 151] => [91, 271] ( 26%, 58% ) | [16, 32, 271] | [16, 32] => [271] ( 21%, 86% ) | [16, 33] | [33] => [16] ( 21%, 79% ) |
| 3 | | [91, 151] => [16, 271] ( 26%, 74% ) | | [32, 271] => [16] ( 21%, 84% ) | [16, 48] | [48] => [16] ( 23%, 80% ) |
| 4 | | [91, 271] => [16, 151] ( 26%, 59% ) | | [61] => [16, 271] ( 22%, 77% ) | [16, 61] | [61] => [16] ( 25%, 88% ) |
| 5 | [16, 91, 151, 271] | [151, 271] => [16, 91] ( 26%, 59% ) | [16, 61, 271] | [16, 61] => [271] ( 22%, 87% ) | [16, 62] | [62] => [16] ( 27%, 79% ) |
| 6 | | [16, 91, 151] => [271] ( 26%, 87% ) | | [61, 271] => [16] ( 22%, 89% ) | | [16] => [91] ( 44%, 58% ) |
| 7 | | [16, 91, 271] => [151] ( 26%, 89% ) | | [62] => [16, 271] ( 22%, 64% ) | [16, 91] | [91] => [16] ( 44%, 83% ) |
| 8 | | [16, 151, 271] => [91] ( 26%, 70% ) | [16, 62, 271] | [16, 62] => [271] ( 22%, 81% ) | [16, 92] | [92] => [16] ( 23%, 73% ) |
| 9 | | [91, 151, 271] => [16] ( 26%, 86% ) | | [62, 271] => [16] ( 22%, 81% ) | [16, 107] | [107] => [16] ( 24%, 82% ) |
| 10 | | | | [91] => [16, 151] ( 29%, 56% ) | [16, 108] | [108] => [16] ( 20%, 78% ) |
| 11 | | | | [151] => [16, 91] ( 29%, 55% ) | [16, 123] | [123] => [16] ( 20%, 79% ) |
| 12 | | | [16, 91, 151] | [16, 91] => [151] ( 29%, 67% ) | [16, 151] | [16] => [151] ( 44%, 59% ) |
| 13 | | | | [16, 151] => [91] ( 29%, 86% ) | | [151] => [16] ( 44%, 82% ) |
| 14 | | | | [91, 151] => [16] ( 29%, 85% ) | [16, 152] | [152] => [16] ( 24%, 72% ) |
| 15 | | | | [91] => [16, 271] ( 37%, 70% ) | [16, 166] | [166] => [16] ( 20%, 88% ) |
| 16 | | | | [271] => [16, 91] ( 37%, 50% ) | [16, 167] | [167] => [16] ( 27%, 80% ) |
| 17 | | | [16, 91, 271] | [16, 91] => [271] ( 37%, 84% ) | [16, 181] | [181] => [16] ( 20%, 91% ) |
| 18 | | | | [16, 271] => [91] ( 37%, 63% ) | [16, 182] | [182] => [16] ( 27%, 79% ) |
| 19 | | | | [91, 271] => [16] ( 37%, 85% ) | [16, 197] | [197] => [16] ( 29%, 80% ) |
| 20 | | | | [286] => [16, 91] ( 20%, 60% ) | [16, 212] | [212] => [16] ( 25%, 51% ) |
| 21 | | | [16, 91, 286] | [16, 286] => [91] ( 20%, 70% ) | [16, 213] | [213] => [16] ( 20%, 76% ) |
| 22 | | | | [91, 286] => [16] ( 20%, 88% ) | [16, 227] | [227] => [16] ( 22%, 86% ) |
| 23 | | | | [107] => [16, 271] ( 20%, 68% ) | [16, 242] | [242] => [16] ( 25%, 82% ) |
| 24 | | | [16, 107, 271] | [16, 107] => [271] ( 20%, 83% ) | [16, 271] | [16] => [271] ( 58%, 78% ) |
| 25 | | | | [107, 271] => [16] ( 20%, 83% ) | | [271] => [16] ( 58%, 79% ) |

⋮

## B.3 $L_1^{LSAR}$ for corresponding Amino acid frequency ranges at site $S_1$

Replace Item No. in Appendix B.2 data with its corresponding Amino acid frequency range as shown in table Appendix A.4.

| Sr. No. | Strong ARs for Frequent 4-itemsets | | Strong ARs for Frequent 3-itemsets | |
|---|---|---|---|---|
| | L | AR(support,confidence) | L | AR(support,confidence) |
| 1 | | <C.0.2> <H.0.2> => { <M.0.2> <W.0.2> } ( 26%, 58% ) | | <D.3.5> => { <C.0.2> <W.0.2> } ( 21%, 72% ) |
| 2 | | <C.0.2> <M.0.2> => { <H.0.2> <W.0.2> } ( 26%, 58% ) | { <C.0.2> <D.3.5> <W.0.2> } | <C.0.2> <D.3.5> => { <W.0.2> } ( 21%, 86% ) |
| 3 | | <H.0.2> <M.0.2> => { <C.0.2> <W.0.2> } ( 26%, 74% ) | | <D.3.5> <W.0.2> => { <C.0.2> } ( 21%, 84% ) |
| 4 | | <H.0.2> <W.0.2> => { <C.0.2> <M.0.2> } ( 26%, 59% ) | | <F.0.2> => { <C.0.2> <W.0.2> } ( 22%, 77% ) |
| 5 | { <C.0.2> <H.0.2> <M.0.2> <W.0.2> } | <M.0.2> <W.0.2> => { <C.0.2> <H.0.2> } ( 26%, 59% ) | { <C.0.2> <F.0.2> <W.0.2> } | <C.0.2> <F.0.2> => { <W.0.2> } ( 22%, 87% ) |
| 6 | | <C.0.2> <H.0.2> <M.0.2> => { <W.0.2> } ( 26%, 87% ) | | <F.0.2> <W.0.2> => { <C.0.2> } ( 22%, 89% ) |
| 7 | | <C.0.2> <H.0.2> <W.0.2> => { <M.0.2> } ( 26%, 89% ) | | <F.3.5> => { <C.0.2> <W.0.2> } ( 22%, 64% ) |
| 8 | | <C.0.2> <M.0.2> <W.0.2> => { <H.0.2> } ( 26%, 70% ) | { <C.0.2> <F.3.5> <W.0.2> } | <C.0.2> <F.3.5> => { <W.0.2> } ( 22%, 81% ) |
| 9 | | <H.0.2> <M.0.2> <W.0.2> => { <C.0.2> } ( 26%, 86% ) | | <F.3.5> <W.0.2> => { <C.0.2> } ( 22%, 81% ) |
| 10 | | | | <H.0.2> => { <C.0.2> <M.0.2> } ( 29%, 56% ) |
| 11 | | | | <M.0.2> => { <C.0.2> <H.0.2> } ( 29%, 55% ) |
| 12 | | | { <C.0.2> <H.0.2> <M.0.2> } | <C.0.2> <H.0.2> => { <M.0.2> } ( 29%, 67% ) |
| 13 | | | | <C.0.2> <M.0.2> => { <H.0.2> } ( 29%, 86% ) |
| 14 | | | | <H.0.2> <M.0.2> => { <C.0.2> } ( 29%, 85% ) |
| 15 | | | | <H.0.2> => { <C.0.2> <W.0.2> } ( 37%, 70% ) |
| 16 | | | | <W.0.2> => { <C.0.2> <H.0.2> } ( 37%, 50% ) |
| 17 | | | { <C.0.2> <H.0.2> <W.0.2> } | <C.0.2> <H.0.2> => { <W.0.2> } ( 37%, 84% ) |
| 18 | | | | <C.0.2> <W.0.2> => { <H.0.2> } ( 37%, 63% ) |
| 19 | | | | <H.0.2> <W.0.2> => { <C.0.2> } ( 37%, 85% ) |
| 20 | | | | <Y.0.2> => { <C.0.2> <H.0.2> } ( 20%, 60% ) |
| 21 | | | { <C.0.2> <H.0.2> <Y.0.2> } | <C.0.2> <Y.0.2> => { <H.0.2> } ( 20%, 70% ) |
| 22 | | | | <H.0.2> <Y.0.2> => { <C.0.2> } ( 20%, 88% ) |

⋮